\documentclass[11pt]{article}%

\usepackage{times}
\usepackage{preamble}
\usepackage{color}

\usepackage{hyperref}

\newcommand{\tv}{\mathrm{tv}}
\newcommand{\gl}{\mathrm{gl}}

\begin{document}

\title{ConvSCCS: convolutional self-controlled case series model for lagged adverse event detection}

\author{
	Maryan Morel$^{\ast,1}$,
	Emmanuel Bacry$^{1,2}$, 
	St\'ephane Ga\"iffas$^{1,3}$, 
	Agathe Guilloux$^{1,4}$, 
	Fanny Leroy$^5$
}

\footnotetext[1]{CMAP Ecole polytechnique 91128 Palaiseau Cedex, France}
\footnotetext[2]{CEREMADE Université Paris-Dauphine, PSL, 75765 Paris Cedex 16, France}
\footnotetext[3]{LPMA Université Paris-Diderot, 75013 Paris, France}
\footnotetext[4]{LaMME Université d'Évry Val d'Essonne, 91037 \'Evry, France}
\footnotetext[5]{Caisse Nationale de l’Assurance Maladie, 75986 Paris Cedex 20, France}

\maketitle

\begin{abstract}
With the increased availability of large databases of electronic health records (EHRs) comes the chance of enhancing health risks screening.
Most post-marketing detections of adverse drug reaction (ADR) rely on physicians' spontaneous reports, leading to under reporting.
To take up this challenge, we develop a scalable model to estimate the effect of
multiple longitudinal features (drug exposures) on a rare longitudinal outcome.
Our procedure is based on a conditional Poisson model also known as self-controlled case series (SCCS).
We model the intensity of outcomes using a convolution between exposures and step functions, that are penalized using a combination of group-Lasso and total-variation.
This approach does not require the specification of precise risk periods, and allows to study in the same model several exposures at the same time.
We illustrate the fact that this approach improves the state-of-the-art for the estimation of the relative risks both on simulations and on a cohort of diabetic patients, extracted from the large French national health insurance database (SNIIRAM), a SQL database
built around medical reimbursements of more than 65 million people.
This work has been done in the context of a research partnership between Ecole Polytechnique and CNAMTS (in charge of SNIIRAM).

\end{abstract}

\section{Introduction}
\label{sec:introduction}

In the recent years, there has been a rapid increase in health data volume and availability.
Large observational databases (LODs) such as claims databases contain electronic health records (EHRs) of millions of patients.
EHRs contain timestamped informations on patients' behavior, such as the drugs purchases, appointments with physicians
or detailed hospitalizations reports.
LODs thus represent an important resource of knowledge for clinical medicine.

One way to leverage this data is adverse drug reaction (ADR) detection.
ADRs are adverse outcomes caused by drugs that have not been detected during prelicensing studies.
There is an important research effort on this matter~\cite{doi:10.1001/jama.2017.5150}.
ADRs are hard to detect, due to their diversity.
They can be related to multiple factors such as dose or time effects or even to patients' susceptibility due to genetic variation, gender, age, etc.~\cite{Aronson1222}.
This paper focuses on time effects, i.e. on the relationship between ADR occurrences and occurrences of other past events (drug purchases).
According to~\cite{Aronson1222}, such effects can be divided into two main categories.
Short-term effects, such as rapid, first dose, or early reactions, occur shortly after the drug exposure.
Long-term effects, such as intermediate, late, or delayed reactions can take months to appear.
One can expect prelicensing studies to catch most of the short terms effects, however, due to their high cost, these studies are generally not run long enough to detect long-term effects.
As a consequence, some ADRs can be identified years after commercialization~\cite{doi:10.1001/jama.2017.5150}.

Detecting long term adverse reactions is thus an important goal of post-marketing surveillance.
Historically, the main effort of post-marketing surveillance is based on spontaneous reports from physicians and consumers~\cite{Schuemie2014}, thus depending on human detection of adverse effects.
Unfortunately, human detection leads to under reporting of ADRs~\cite{Alvarez-Requejo1998}.
Spontaneous reports often trigger statistical studies using LODs.
Studies such as~\cite{Neumann2012, Basson2015} or~\cite{Maura2015} illustrate the usefulness of LODs for drug safety studies.
Using LODs more extensively could improve ADR detection by generating hypotheses directly from the data using screening strategies~\cite{Trifiro2009}.
In the recent years, this perspective led to an increased research effort involving the use of LODs~\cite{Hripcsak2015}.

However, using LODs for ADR screening is not a trivial task.
A first challenge comes from the scale of the data.
Indeed, EHRs data allows to study millions of patients across several years, hence it requires the use of scalable algorithms.
A second challenge comes from the fact that LODs are generally administrative databases (commonly designed for accounting purposes in the context of health insurance), and not designed for medical purposes.
As a result, they often do not include confounding clinical variables, which can lead to seriously biased results.
A last challenge results from the number of drugs the patients are exposed to.
When using LODs for risk screening, a priori knowledge on the potentially problematic drugs might be scarse, consequently the number of combination of drugs and outcomes to consider is potentially very large.

Self-Controlled Case Series (SCCS) models, originally developed for vaccine safety studies~\cite{Farrington1995} can help for post-marketing studies using LODs~\cite{Gault2017}.
A SCCS model scales quite well since it is fitted on cases only, i.e. patients with at least one ADR.
Moreover, as explained below, its goodness-of-fit function cancels out non-longitudinal confounders, which reduces potential non-longitudinal biases.
Thus, a SCCS model helps with the first two challenges described previously. However, a SCCS model relies heavily on the definition of a time-at-risk period, which makes it hard to use in multivariate settings, such as the study of the impact of several longitudinal exposures to drugs on the occurrence of an ADR.

This paper introduces a new approach in the framework of SCCS models that addresses the three challenges mentioned previously.
It considers in a flexible way several longitudinal features at the same time (longitudinal drug exposures), and the time occurrences of an ADR as the outcome.
This models learns automatically potentially fast or delayed effects of these exposures on the outcome, with no precise knowledge on a time-at-risk period.
It therefore provides an important extension to the usage of SCCS models, allowing to \emph{study multiple exposures at the same time, while requiring much less attention to the definition of time-at-risk periods}.

Previous attempts to solve this problem relied on the use of splines to provide a more flexible modeling of drug effects \cite{Schuemie2014,Ghebremichael-Weldeselassie2016,Ghebremichael-Weldeselassie2017}. 
However, the use of splines makes the estimation of the model more complicated, resulting in models able to fit the effect of a single drug in addition to a temporal baseline. 
This can be problematic when performing ADRs screening, as SCCS is sensitive to temporal confounders, and thus, to the omission of longitudinal features.
In a more recent work,~\cite{Bao2017} use step functions within a multiple drugs setting. However, their work is no longer an SCCS approach since full-likelihood estimation is used though only cases are kept.

The paper is organized as follows.
We first describe SCCS models in Section~2 and construct our method in Section~3.
Numerical experiments are given in Section~4.
It includes in Section~\ref{sub:simulations} experiments on simulated data, with a comparison to state-of-the-art methods from the SCCS literature. Simulations are done in a realistic way, for the application to the LOD considered in the paper.
Section~\ref{sub:cnam-application} gives an application of our method on a LOD from the French national health insurance information system  (SNIIRAM, an SQL database
built around medical reimbursements of more than 65 million people). It illustrates the ability of the method to reveal the known effect (previously identified by \cite{Neumann2012}) of glucose-lowering drugs on the risk of bladder cancer.
A conclusion is given in Section~5, discussing the limitations and possible extensions of this work, and mathematical and numerical details are provided in an Appendix.

\section{Self-controlled case series models}
\label{sec:model}

SCCS models allow to estimate the impact of longitudinal features (such as time-varying exposures to drugs) on the occurrence intensity of events of interest (such as dates of adverse events), see~\cite{Farrington1995}.
An interesting particularity with this family of methods is that individuals form their own controls: individuals who do not experience the event of interest are not used to fit the model.
This construction relies on the property of order statistics of the Poisson process
and the statistical output of such models is an estimation of the
\emph{relative incidence} of the longitudinal features, i.e. the relative increase of the intensity of outcomes.

\subsection{Conditional Poisson regression and SCCS models}
\label{sub:conditional_poisson}

Data is available from a global observation period $(a,b]$, where the time can be either calendar, or measured by the age of individuals.
Each patient $i=1, \ldots, m$ has an observation period $(a_i, b_i] \subset (a, b]$, in which we observe:
\begin{itemize}
	\item the time occurrences $t_{i, 1} < \cdots < t_{i, n_i}$ of the event of interest (also called \emph{outcome} in what follows), where $n_i$ is the total number of outcomes of $i$; 
	\item a vector of $d$ longitudinal features
	\begin{equation*}
		X_i = \big(X_i(t) =(X^1_i(t) \; \cdots \; X^d_i(t)) : t \in (a_i, b_i]\big),
	\end{equation*}
	where in the context of drug safety studies, $X_i^j(t)$ gives us information about the exposure of patient $i$ to drug $j$ at time $t \in (a, b]$.
\end{itemize}
To the outcomes of $i$ on $(a_i, b_i]$ can be associated a counting process $N_i$ (by defining $N_i(t) = \sum_{k=1}^{n_i} \ind{t_{i, k} \leq t}$). There are three main assumptions in SCCS models~\cite{Farrington2006a}:
\begin{itemize}
	\item[(1.)]  the features are considered to be exogenous, meaning that the counting process $N_i$ does not have any influence on the features $X_i$. This allows the conditioning of the full trajectory $X_i$ in~\eqref{eq:intensity-general-definition};
	\item[(2.)] the interval of observation $(a_i,b_i]$ is supposed to be independent of $N_i$ and all the following computations have to be understood conditionally to it;
	\item[(3.)] $N_i$ is assumed to be a Poisson process conditionally to $X_i$.
\end{itemize}
As a consequence, its conditional intensity is given by
\begin{equation}
	\label{eq:intensity-general-definition}
	\lambda_i(t, X_i) = \P(dN_i(t)=1~|~ X_i)
\end{equation}
for $t \in (a_i,b_i]$.
This model can be therefore understood as a regression model, allowing to regress the outcomes in $N_i$ on the longitudinal features $X_i$.
Also note that $n_i = \int_{(a_i, b_i]} d N_i(t)$.

The idea of the SCCS method is to condition on both $X_i$ and $n_i$. Usual arguments (see Section~\ref{supp:likelihood_sccs}) imply that the likelihood of $N_i | (X_i, n_i)$ of $i=1, \ldots, m$ independent patients is proportional to
\begin{equation}
\label{eq:sccs-likelihood}
 \prod_{i = 1}^m \prod_{k=1}^{n_i} \frac{ \lambda_i(t_{i, k}, X_i)}{\int_{a_i}^{b_i} \lambda_i(s,X_i) ds}.
\end{equation}
Note that the conditioning with respect to $n_i$ induced two notable properties of~\eqref{eq:sccs-likelihood}:
\begin{itemize}
  \item[(i)] it only depends on patients $i$ such that $n_i \geq 1$ (while the ``full'' likelihood of $N_i | X_i$ does depend on patients $i$ for whom $n_i = 0$);
 	\item[(ii)] it involves ratios of intensities only, leading to a likelihood that does not depend on time-invariant parameters when using a multiplicative parametric modeling of the intensity.
 \end{itemize}
An interesting consequence of~(i) is an improved scalability: the number of individuals $i$ to consider is reduced to the ones with $n_i \geq 1$, which is beneficial when studying rare adverse effects in large LODs.
Point~(ii) entails that the estimation method ignores non-temporal effects, by estimating only the relative incidences.
This makes in particular SCCS models particularly robust to the patient's susceptibility.
These two properties are particularly appealing in our case, as it allows to handle the volume of LODs and the missing informations of EHRs contained in claims databases.

In order to study acute vaccine adverse effects,~\cite{Farrington2006a} considers the following model for the intensity:
\begin{equation*}
	\lambda(t, X_i) = \exp \big( \psi_i + \gamma_i + \phi(t) + X_i(t)^\top \beta \big),
\end{equation*}
where $\psi_i$ is the baseline incidence of patient $i$ and $\gamma_i$ is a sum of non-temporal fixed and random individual effects.
The parameter $\phi(t)$ is a time-dependent baseline which is common to all individuals.
If age is used as the time scale, this term can help to capture age effects.
The vector of parameters $\beta \in \R^d$ quantifies the effect of the longitudinal features $X_i(t)$ on the intensity.
Because of Point~(ii) above, the non-longitudinal effects $\psi_i$ and $\gamma_i$ cancel out in the likelihood~\eqref{eq:sccs-likelihood}, which makes the model robust to non-longitudinal confounders. Relative incidences can easily be computed by taking the exponential of the corresponding coefficient, such as $\exp(\phi(t))$ for the baseline relative incidence.

SCCS models were initially designed for drug safety studies~\cite{Farrington1995}, using the suspected ADR as the outcome. In this context, estimating the relative incidence of drug use requires to define time-at-risk periods related to drug use in which the suspected ADR might occur. The longitudinal features $X_i(t)$ are then used to express the fact of being at risk or not for a particular drug, at time $t$.
One must then determine for how long patients are at risk after each exposure to a drug, and if this risk occurs either immediately or after some amount of time.
Defining proper time-at-risk windows is a hard problem when studying a single (drug, ADR) pair, which worsens even further when considering a set $(\mathrm{drug}_1, \mathrm{ADR}), \ldots, (\mathrm{drug}_d, \mathrm{ADR})$ of such pairs.
In the case of ADR screening over multiple drugs, such a methodology might even become inappropriate.

\subsection{Risk screening} 
\label{sub:risk_screening}

When prior knowledge on time-at-risk window is not available, a simple method is to use a large window in order to be sure to capture the potential effect.
However, this strategy typically ``dilutes'' the risk over the window, see~\cite{Xu2011}, leading to a model unable to detect ADRs.
Existing works propose to relax the time-at-risk window definition while trying to overcome this risk dilution. It is proposed in~\cite{Xu2011, Xu2013} to select an optimal risk window by testing several window sizes, in a data-driven fashion.
However, this method is difficult to adapt for ADR screening when considering $d \geq 1$ drugs and $q$ risk windows at the same time, since it requires to fit $q^d$ models.

A different approach relies on fitting time-dependent parameters in order to estimate the risk of ADR over large risk windows.
The model estimates a time-varying relative incidence function all along the risk window instead of assuming it to be constant.
This approach is used in~\cite{Schuemie2014}, where the drug effect is a function $\theta$ of the accumulated exposures.
It uses a discrete model with daily granularity, assuming that the integral of $X_i(t)$ over one day is equal to $1$ when the patient is exposed to the studied drug.
Accumulated exposures up to time $t$ is measured by $\int_{a_i}^t X_i(s) \text{d} s$, where $X_i(t)$ is univariate, and expresses the exposure to a single drug at time $t$, leading to the following model for the intensity:
\begin{equation*}
  \lambda_i(t, X_i) = \exp \Big(\psi_i + \gamma_i + \phi(t) + \theta\Big( \int_{a_i}^t X_i(s) \text{d} s \Big)
  + X_i(t) \beta \Big),
\end{equation*}
where the function $\theta$ is estimated using natural cubic splines.
As the splines are not regularized, this model might be prone to overfitting.
Alternatively,~\cite{Ghebremichael-Weldeselassie2016} use a convolution to model drug effects, writing the intensity as
\begin{equation*}
  \lambda_i(t, X_i) = \exp \big(\psi_i + \gamma_i + \phi(t)\big)  \int_{a_i}^t X_i(s) \theta(t - s)\text{d} s.
\end{equation*}
In this model, $X_i(t)$ is either a point exposure $X_i(t) = \delta_{c_i}(t)$ where $\delta_{c_i}$ stands for a Dirac mass at date $c_i \in \R^+$, or a continuous exposure to a constant quantity $x$, namely $X_i(t) = x\ind{(c_{i}, b_{i}]}(t)$.
In the former case, the intensity can be expressed as
\begin{equation}
	\label{eq:gw-intensity}
  \lambda_i(t, X_i) = \exp \left(\psi_i + \gamma_i + \phi(t)\right)  \theta(t - c_i).
\end{equation}
The weighting function $\theta$ is estimated using M-splines (in order ensure positivity) in~\cite{Ghebremichael-Weldeselassie2016,Ghebremichael-Weldeselassie2017}, while the age effect $\phi$ is estimated by step functions
in~\cite{Ghebremichael-Weldeselassie2016} and by splines in~\cite{Ghebremichael-Weldeselassie2017}.
The model considered in~\cite{Ghebremichael-Weldeselassie2016,Ghebremichael-Weldeselassie2017} could deal with multiple point exposures $c_i$ for the drug, given that the maximum time gap between successive exposures is smaller than the support of $\theta$, but the authors did not developed this point.
The parameter $p$ can be chosen loosely, it just needs to be large enough to capture the risk increase if this risk is delayed or late. If $p$ is too large, the formulations presented above should still be able to provide good risk estimates. Selecting $p$ is then much simpler than designing a time-at-risk window.

Both~\cite{Schuemie2014} and~\cite{Ghebremichael-Weldeselassie2016, Ghebremichael-Weldeselassie2017} seem restricted to
the study of a single (drug, ADR) pair at a time.
This can be problematic since SCCS is sensitive to time-varying confounders and benefits from studying multiple drugs at once as shown by both \cite{Simpson2013a} and
\cite{Moghaddass2016}.
In order to fit a SCCS model using several drugs at the same time,~\cite{Ghebremichael-Weldeselassie2017} propose to extend their work by modeling additional drugs effect with step functions instead of splines.
However, such functions are basically not regularized, which can result in overfitting, and are very sensible to the number of steps used.
We finally mention the recent work of~\cite{Bao2017} which develop a similar approach, using step functions in a multiple drugs setting. In this work, the step functions weights express the risk associated with time intervals partitioning time-at-risk windows. However, let us point out that this model is no longer, strictly speaking, an SCCS approach since, though non-cases are filtered out, full likelihood estimation is performed.
We believe that this filtering, with no correction of the likelihood, could lead to an estimation bias.


\section{ConvSCCS: an extension of SCCS models}
\label{sec:convsccs}

We introduce ConvSCCS, an extension for the SCCS model, that allows to incorporate exposures to several drugs at the same time. Our model is time-invariant thanks to a convolutional structure, and estimate easily interpretable relative incidence curves, thus enabling multivariate lagged effect detection.
More specifically, we construct a model that estimates the effect of longitudinal features using step functions convolved with point drug exposures.
In this model, the number of steps which is required to detect effects in a precise manner is large, leading to an over-parametrized model with poor estimation accuracy.
We solve this issue in Section~\ref{sec:estimation} below by using a penalization technique which is particularly relevant in this context.
It combines total-variation and Group-Lasso penalizations, in order to obtain estimations that are, whenever statistically relevant, piecewise constant over larger steps, therefore reducing the actual number of parameters in the model.
As illustrated in Section~\ref{sec:experiments}, this leads to improvements of current state-of-the-art methods, and provides interpretable results on the observational database considered in this paper, see in particular
Section~\ref{sub:cnam-application}.

\subsection{Discrete convolutional SCCS}
\label{sub:discrete_convolutional_sccs}

We assume that, for $i=1, \ldots, m$, the intensity $\lambda$ is constant over time intervals
$I_k = (t_k,t_{k+1}]$, $k=1, \dots, K$ that form a partition of the observation interval $(a, b]$.
We choose $I_k$ to be of constant length, chosen without loss of generality equal to $1$.
In practice, we use the smallest granularity allowed by data.
We therefore can assume that $(a_i, b_i]\cap I_k$ is either $\emptyset$ or $I_k$ for all $i=1, \ldots, m$, and $k=1, \dots, K$, which means that the observation period of each individual is a union of intervals~$I_k$.
Denoting by $\lambda_{i, k}$ the value of $\lambda(t, X_i(t))$ for $t \in I_k$, and defining $y_{ik} :=  N_i(I_k)$, the discrete SCCS likelihood can be written as
\begin{equation*}
	L(y_{i1}, \dots, y_{ik} | n_i, X_i) = n_i ! \prod_{k=1}^K \left( \frac{ \lambda_{ik}}{\sum_{k'=1}^K
	\lambda_{ik'}} \right) ^{y_{ik}},
\end{equation*}
where we use the convention $0^0 = 1$, i.e. only the exposition period $(a_i,b_i]$ contributes to the likelihood, and since $N_i(I_k) = \lambda_{ik} = 0$ whenever $I_k \cap (a_i,b_i] = \emptyset$, see Section~\ref{supp:discrete_sccs} for more details.
We consider an intensity given by
\begin{equation*}
  \lambda_i(t, X_i) = \exp \Big(\psi_i + \gamma_i + \phi(t) + \int_{a_i}^t
  X_i(s)^\top \theta(t - s)\text{d} s \Big).
\end{equation*}
Since the intensity is constant on each $I_k$, it can be rewritten as
\begin{equation*}
  \lambda_{ik}(X_i) = \exp \Big(\psi_i + \gamma_i + \phi_k + \sum_{k'=a_i}^k
  X_{ik'}^\top \theta_{k - k'} \Big),
\end{equation*}
where $X_{ik}$ stands for the value of $X_i(t)$ for $t \in I_k$ and $\theta \in \R^{d \times K}$.
We observe $l=1, \dots, L_i^j$ starting dates of exposures $c_{il}^j$ such that $\min_{l, l'} |c_{il}^j - c_{il'}^j| > p$ to avoid overlaps between exposures, and introduce the features $X^j_{ik} = \sum_{l=1}^{L_i^j} \ind{k=c_{il}^j}$, which leads to the following intensity
\begin{equation}
	\label{eq:our_intensity}
	\lambda_{ik}(X_i) = \exp\Big(\psi_i + \gamma_i + \phi_k + \sum_{j=1}^d
	\sum_{l=1}^{L_i^j} \theta_{k-c_{il}^j}^j \ind{[0,p]}(k-c_{il}^j) \Big).
\end{equation}
The quantity $\exp(\theta_k^j)$ corresponds to the relative incidence of an exposure to drug $j$ that occurs~$k$ time units after an exposure start.
Finally, the likelihood is equal to
\begin{equation}
\label{eq:our_likelihood}
L(y_{i1}, \dots, y_{ik} | n_i, X_i) = \prod_{k=1}^K \left( \frac{ \exp\big( \phi_k + \sum_{j=1}^d \sum_{l=1}^{L_i^j} \theta_{k-c_{il}^j}^j \ind{[0,p]}(k - c_{il}^j) \big)}{\sum_{k'=1}^K \exp\big( \phi_{k'} + \sum_{j=1}^d \sum_{l=1}^{L_i^j}
\theta_{k' - c_{il}^j}^j \ind{[0, p]}(k' - c_{il}^j) \big) } \right)^{y_{ik}}
\end{equation}
and depends only on the parameters $\theta$ for the exposures and the age effects $\phi$.
\label{relative_incidence}

\subsection{Penalized estimation} 
\label{sec:estimation}

This formulation of intensity~\eqref{eq:our_intensity} is flexible since it allows to capture an immediate effect in $\theta_0^j$, or delayed ones using $\theta_k^j$ for $k \geq 1$.
This flexibility comes at a cost: it increases significantly the number of parameters to be estimated, which might lead to inaccurate estimations and to overfitting of the dataset.
We therefore introduce a penalization technique which allows to handle this issue, and which provides interpretable estimations of the relative risks as a byproduct.

We introduce groups $\theta^j = [\theta_1^j \cdots \theta_p^j] \in \R^p$ of parameters that quantify the impact of exposures to drugs $j=1, \ldots, d$ at different lags $k=1, \ldots p$.
We want to induce two properties on the relative risks of the drugs: a ``smoothness'' property along lags $k=1, \ldots, p$, namely we want consecutive relative risks $\exp(\theta_k^j)$ and $\exp(\theta_{k-1}^j)$ to be basically close; and
the possibility for a drug to have no effect, namely to induce that $\theta^j$ can be the null vector.
This can be achieved with the following penalization that combines total-variation and group-Lasso
\begin{equation}
\label{eq:our_penalization}
	\pen(\theta) = \gamma_{\tv} \sum_{j=1}^J \sum_{k=1}^{p-1} | \theta^j_{k+1} - \theta^j_{k} | +
	\gamma_{\gl} \sum_{j=1}^J \| \theta^j \|_2
\end{equation}
over the groups $\theta^j$ for $j=1, \ldots, d$, where $\gamma_{\tv} \geq 0$ and $\gamma_{\gl} \geq 0$
are respectively levels of penalization for the total-variation and the group-Lasso.
Total-variation is a convex proxy for segmentation, which is known as \emph{fused-lasso},
see~\cite{TibRosZhuKni-05}, whenever used with an extra $\ell_1$-penalization.
It is very popular for image denoising and deblurring, see for
instance~\cite{chambolle2010introduction}. The group-Lasso introduced in~\cite{yuan2006model} acts like the lasso at the group level: depending on $\gamma_{\gl}$, it can cancel out a full block $\theta^j$.
Total-variation penalization is also known to consistently estimate change points for the estimation of the intensity of a Poisson process, see~\cite{Alaya2015}.

We therefore write the penalized negative log-likelihood of our model as follows:
\begin{equation}
\label{eq:penalized-log-lik}
- \ell(\phi, \theta) + \pen(\theta) = - \frac 1n \sum_{i=1}^n \sum_{k=1}^K y_{ik}
 \log \Bigg( \frac{\lambda_{ik}(\phi, \theta)}{\sum_{k'=1}^K \lambda_{ik}(\phi, \theta)} \Bigg) + \pen(\theta),
\end{equation}
where $\pen$ is given by~\eqref{eq:our_penalization} and where we recall that
\begin{equation*}
	\lambda_{ik}(\phi, \theta) = \exp\Big(\phi_k + \sum_{j=1}^d
	\sum_{l=1}^{L_i^j} \theta_{k-c_{il}^j}^j \ind{[0,p]}(k-c_{il}^j) \Big).
\end{equation*}
The function~\eqref{eq:penalized-log-lik} is convex and $\ell(\phi, \theta)$ is gradient-Lipschitz.
However, the sparsity-inducing penalization $\pen(\theta)$ is not differentiable: we therefore use a proximal
first order method to minimize efficiently~\eqref{eq:penalized-log-lik}.
Namely, we use the state-of-the-art SVRG algorithm from~\cite{Xiao2014}, which is a fast stochastic proximal gradient descent algorithm, using a principle of variance reduction of the stochastic gradients.

Finally, the hyper-parameters $\gamma_{\tv}$ and $\gamma_{\gl}$ are selected using a stratified V-Fold cross-validation on the negative log-likelihood.
The validation error is estimated by a simple average over the test folds, and we select the best hyper-parameters using the ``one-standard-error'' rule from~\cite{Friedman2010}.

\section{Experiments}
\label{sec:experiments}

In this Section we compare ConvSCCS with state-of-the-art baselines, namely
SmoothSCCS~\cite{Ghebremichael-Weldeselassie2016} and
NonparaSCCS~\cite{Ghebremichael-Weldeselassie2017}, that are described below, see also~Section~\ref{sub:risk_screening} for further details.

\emph{ConvSCCS} is the method introduced in this paper: an extension of SCCS models allowing to fit the effect of \emph{several} drugs on a ADR in a flexible way, see also Table~\ref{tab:algo-comparison} below.
ConvSCCS is available in our open-source \texttt{tick} library, see Section~\ref{supp:implementation} for details.

\emph{SmoothSCCS} is introduced
in~\cite{Ghebremichael-Weldeselassie2016}, which
uses a convolution to model the effect of a \emph{single} drug exposure to a disease and step functions to model the effect of age. We use the \texttt{SCCS} R package implementation, available at \url{http://statistics.open.ac.uk/sccs/r.htm}. We use 12 knots and six groups of age as suggested
in~\cite{Ghebremichael-Weldeselassie2016}.
Since this model is designed to fit (drug, ADR) pairs, we fit the model on each drug successively.

\emph{NonparaSCCS} is introduced in~\cite{Ghebremichael-Weldeselassie2017}
which uses splines to model both the effect of drug exposure and age. We use the same R package and settings as the ones described for SmoothSCCS.

We did not include~\cite{Schuemie2014} as we have not found any open source implementation of this work. We did not tried to use~\cite{Simpson2013a} since we do not have precise priors on relevant risk periods in the context of ADR screening.

\begin{table}[!h]
\begin{tabular}{rcccc}
\hline
Algorithm & Regularized & Multiple features & Multiple exposures & Flexible effect\\
\hline

MSCCS & yes & yes & yes & no \\
ESCCS & no  & no  & accumulated & yes\\
SmoothSCCS & yes & no  & no  & yes\\
NonParaSCCS & yes & no  & no  & yes\\
ConvSCCS & yes & yes & yes & yes\\

\hline
\end{tabular}
\caption{Comparison of SCCS methods with ConvSCCS.
MSCCS is introduced in~\cite{Simpson2013}, ESCCS in~\cite{Schuemie2014}, while SmoothSCCS and NonParaSCCS are respectively introduced in ~\cite{Ghebremichael-Weldeselassie2016,Ghebremichael-Weldeselassie2017}.
Regularized models are constrained to avoid overfitting, the constraint being controlled by hyper-parameters.
The models can either fit multiple features at a time, or be limited to study only one feature at a time. We do not consider SmoothSCCS and NonParaSCCS as able to study multiple features properly, since only one feature can be regularized.}
\label{tab:algo-comparison}
\end{table}

\subsection{Simulations}
\label{sub:simulations}

The performances of our model against SmoothSCCS and NonParaSCCS are compared through a simulation study.
For this purpose, multivariate longitudinal exposures and outcomes are simulated, with a correlation structure between exposures that mimic typical datasets available in large observational databases (LODs).

\paragraph{Simulation of longitudinal features.}

The simulation of realistic longitudinal features is a difficult task, for which we use Hawkes processes, see~\cite{hawkes1974cluster}, which is a family of counting process with an autoregressive intensity.
An example of simulated features for an individual is displayed in Figure~\ref{hawkes_adjacency} below.
Namely, we simulate dates of purchases $\{t_i^j\}_{i \geq 1}$, of drugs $j=1, \ldots, d$ using a multivariate Hawkes process $N_t = [N_t^1 \cdots N_t^d]$, for $t \geq 0$,
where $N_t^j = \sum_{k \geq 1} \ind{t_k^j \leq t}$ for any $t \geq 0$.
The process $N_t$ is a multivariate counting process, whose components $N^j$ have intensities
\begin{equation}
	\label{eq:hawkes-intensity}
	\lambda_t^j = \mu_j + \sum_{j'=1}^d \sum_{k \geq 1} A_{j, j'}
	\alpha \exp(-\alpha (t - t_k^{j'}))
\end{equation}
for $j=1, \ldots, d$. This corresponds to a Hawkes process with so-called \emph{exponential kernels}.
The $\mu_j \geq 0$ are called \emph{baselines} intensities, and correspond to the exogenous probability of being exposed to drug $j$.
In the matrix $A = [A_{j, j'}]_{1 \leq j, j' \leq d}$, called the \emph{adjacency matrix}, the entry $A_{j, j'} \geq 0$ quantifies the impact of past exposures to drug $j'$ on the exposition intensity
to drug $j$ and $\alpha > 0$ is a memory parameter.
\begin{figure}[!h]
  \centering
  \includegraphics[width=0.9\textwidth]{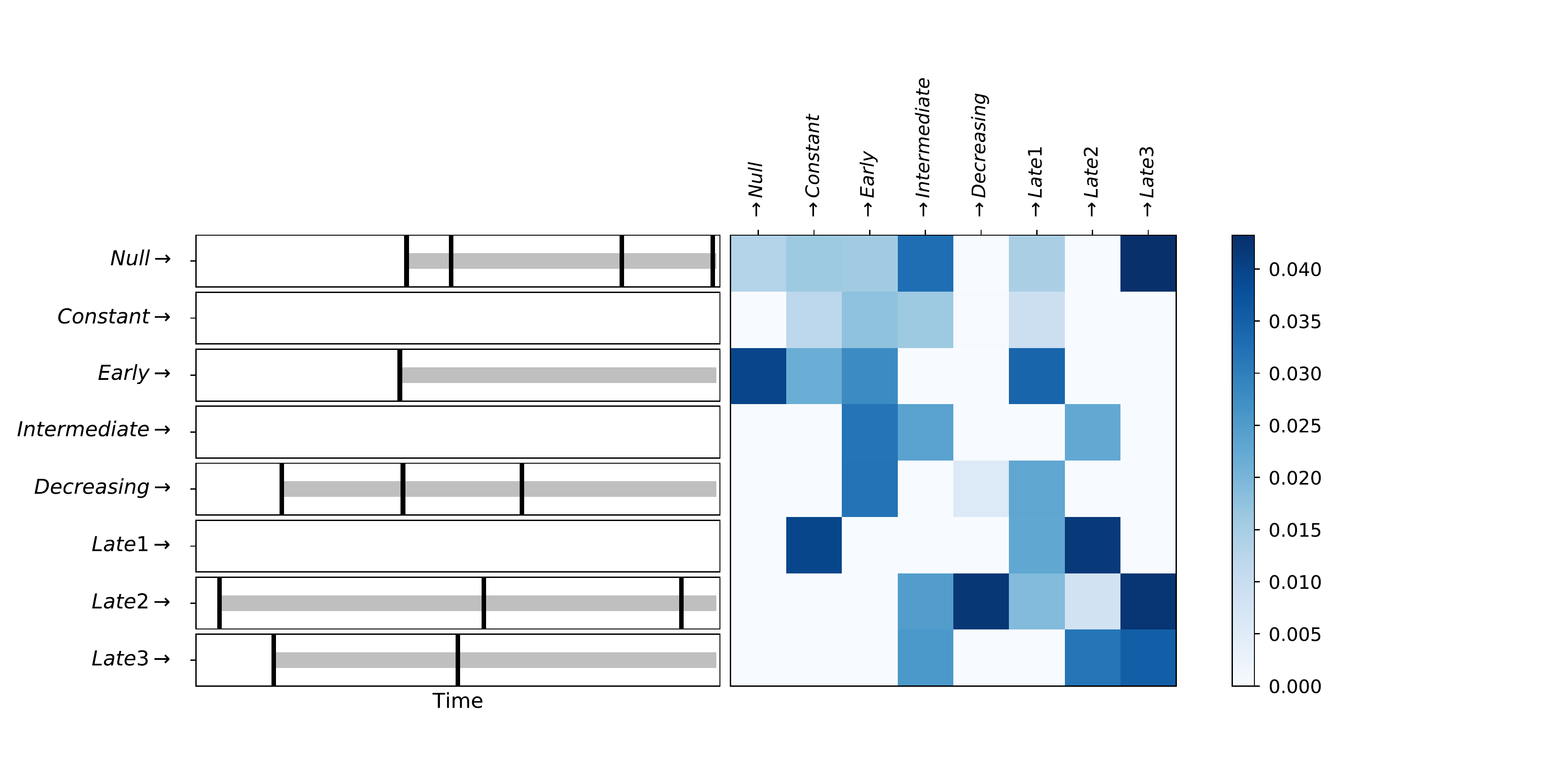}%
  \caption{
  \emph{Left}: example of simulated dates of drugs purchases (vertical black lines). Exposure starts at the date of the first purchase (gray horizontal lines).
	\emph{Right}: example of generated adjacency matrix $A$ for longitudinal feature simulation using the Hawkes process. This matrix encodes the correlation structure of exposures to drugs. To ease the reading, this figures represent the transposed adjacency matrix $A^\top$. For example, a purchase of a `null' drug increases the probability of purchasing a `Late3' drug.
	In \emph{Left} and \emph{Right} we simulate potential exposures to 8 drugs, each of them have a different risk profile (named ``null'', ``constant'', ``early'', etc.). These profiles are described in Section~\ref{sub:simulation_details}.
	}\label{hawkes_adjacency}
\end{figure}
Our simulation setting, detailed in Section~\ref{sub:simulation_details}, has been chosen so that it generates exposures that resemble as much as possible to actual exposures from the LOD considered in this paper.
A single matrix $A$ is simulated for the whole population, but a new one is generated in each round of simulation.
Recalling that the simulated events $t_i^j$ correspond to the purchase date of drugs (this is the only information available in the LOD described in Section~\ref{sub:cnam-application} below), we consider that a patient is exposed to a drug $j$ at time $t_1^j$.

\paragraph*{Simulation of relative risks.}

We assume that all simulated adverse outcomes can take place at most 50 time intervals after the first exposure.
We consider two sets of relative risk profiles from~\cite{Ghebremichael-Weldeselassie2017} and~\cite{Aronson1222}. These sets are precisely described in Section~\ref{sub:simulation_details}, and contain several types and shapes of risks profiles.

\paragraph{Simulation of outcomes.}

We simulate $m=4000$ patients' exposures over $K=750$ time intervals.
To allow individuals to have observation periods of different lengths, we sample $b_i$ as $B - e_i$ where $e_i$ are sampling using an exponential distribution with intensity $1 / 250$, so that on average, the end of observation occurs around the $500th$ time interval.
Once the exposures are simulated, we compute intensities $\lambda_{ik}$ for each individual using simulated exposures and the risk profiles described above.
Intensities $\Lambda_{ik}$ are set to zero for all $k > b_i$.
Without loss of generality, we consider each case to have experienced only one event.
The cases can be simulated easily using a multinomial distribution $\mathrm{Mult}(1 ; p_{i, 0}, \dots, p_{iK})$ where $p_{ik} = \lambda_{ik} / \sum_{k'=1}^K\mathbf{\lambda}_{ik'}$.

\paragraph{Performance measure.}

\label{MAE} The performance of the different models is computed using the mean absolute error (MAE) between the estimated relative incidence and the true risk profile, see Section~\ref{sub:simulation_details} for details. 
For both sets of relative risk profiles, we simulate $m = 4000$ cases, and simulate $100$ datasets for each scenario.

\paragraph{Results.}

\begin{figure}[!h]
\centering
\includegraphics[width=.8\textwidth]{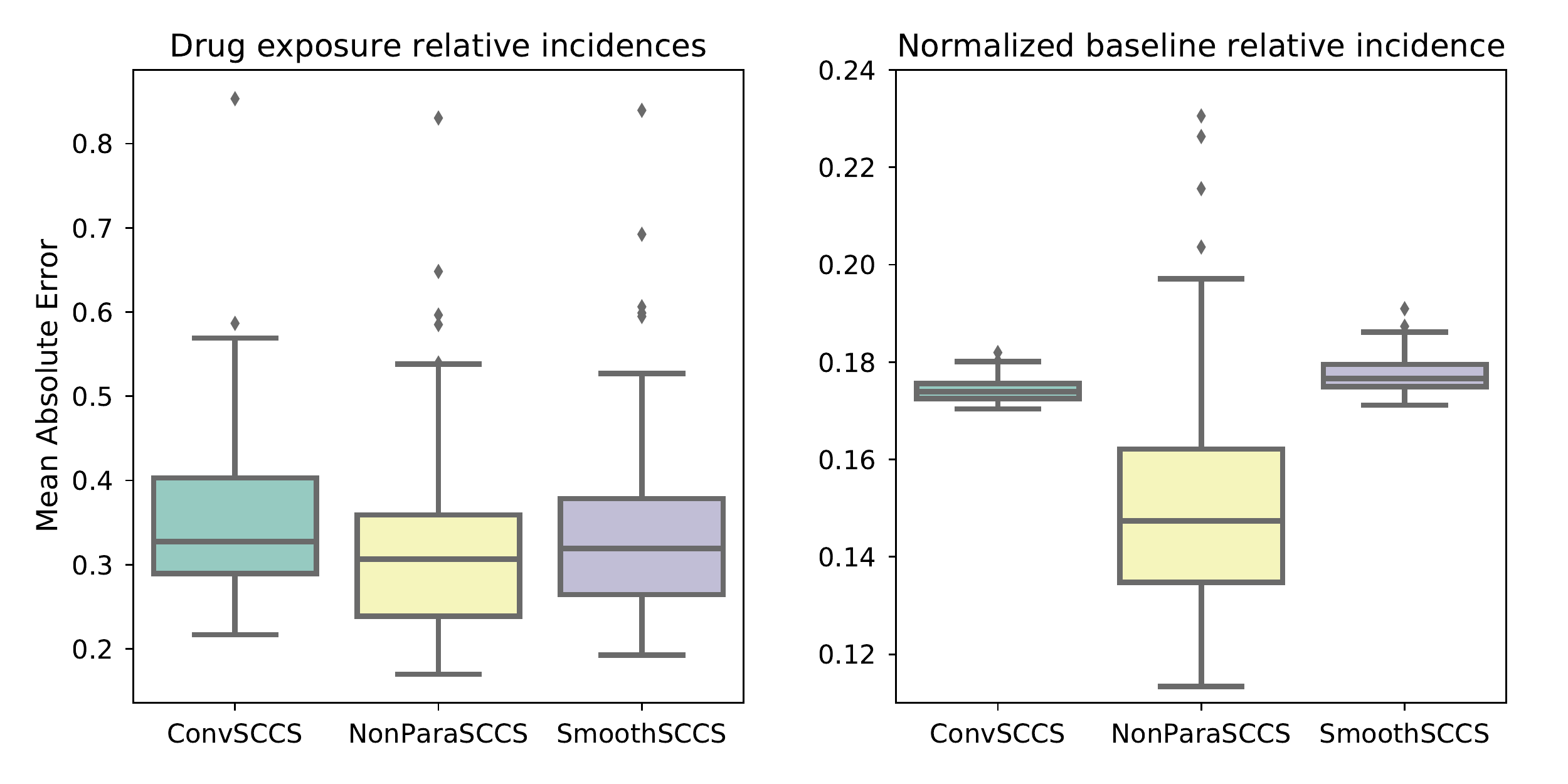}
\caption{
Simulations results using Set~1 or risk profiles (see Figure \ref{sim_effects_1}) with $m=4000$. The boxplots represent the distribution of mean absolute error as defined in Section~\ref{MAE}, computed over $100$ simulated populations. \emph{Left:} MAE distribution of the drug exposure relative incidences. \emph{Right:} MAE distribution of the baseline relative incidences, constrained so that their integral is equal to one.}
\label{simulations_results_set1}
\end{figure}

\begin{figure}[!h]
\centering
\includegraphics[width=.8\textwidth]{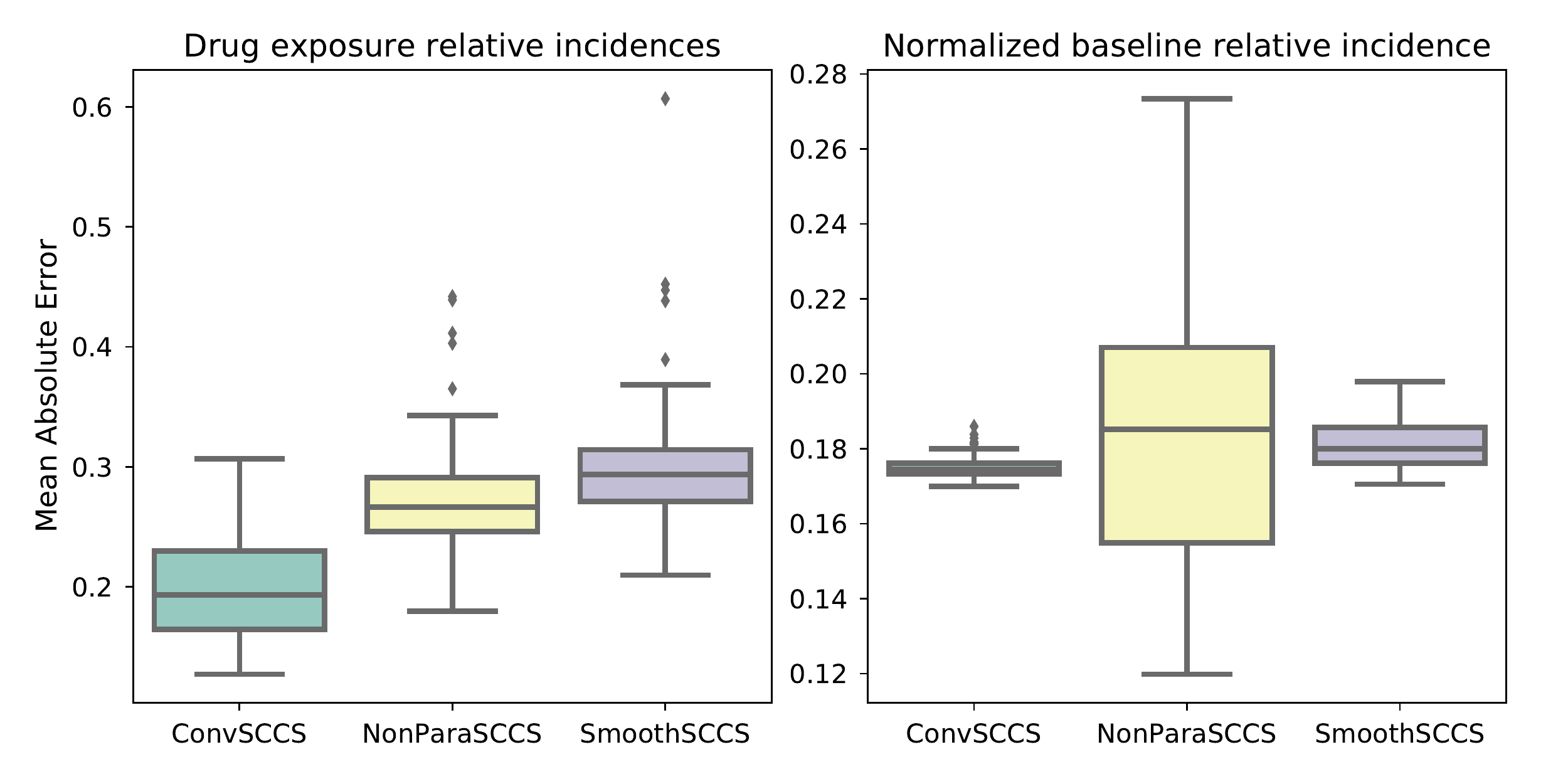}
\caption{
Simulations results using Set~2 or risk profiles (see Figure \ref{sim_effects_2}) with $m=4000$. The boxplots represent the distribution of mean absolute error as defined in Section~\ref{MAE}, computed over $100$ simulated populations. \emph{Left:} MAE distribution of the drug exposure relative incidences. \emph{Right:} MAE distribution of the baseline relative incidences, constrained so that their integral is equal to one.}
\label{simulations_results_set2}
\end{figure}

Boxplots representing the MAE distribution over the $100$ simulated datasets are represented
in Figures~\ref{simulations_results_set1} and~\ref{simulations_results_set2}.

In Set~1 of relative exposures, which is an ``easy'' setting ($4$ features and $8$ non-zero elements in matrix $A$, see Section~\ref{sub:simulation_details}), the gain resulting from studying several drugs at a time seems to be balanced by the bias resulting from using step functions when fitting smooth risk profiles.
Indeed, as shown by Figure~\ref{simulations_results_set1}, the errors of estimation of the relative risks of exposures to drugs are similar across the three considered models.
For the estimation of the baseline NonParaSCCS performs better than ConvSCCS and SmoothSCCS, since the use of splines results in a better approximation than the step functions with six groups of age.

In Set~2 of relative exposures, which is a more difficult setting ($14$ features, with $24$ non-zero elements in~$A$, see Section~\ref{sub:simulation_details}), ConvSCCS outperforms both SmoothSCCS and NonParaSCCS.
We observe in Figure~\ref{simulations_results_set2} that fitting the effect of several drugs at the same time and using our penalization provides a better estimation accuracy than NonParaSCCS and SmoothSCCS, the improvement being larger for the estimation of the risks profiles of drugs exposures than for the baseline.
This illustrates the benefits of fitting several drugs at the same time in the context of a SCCS model.

\begin{figure}[!h]
  \centering
  \includegraphics[width=.8\textwidth]{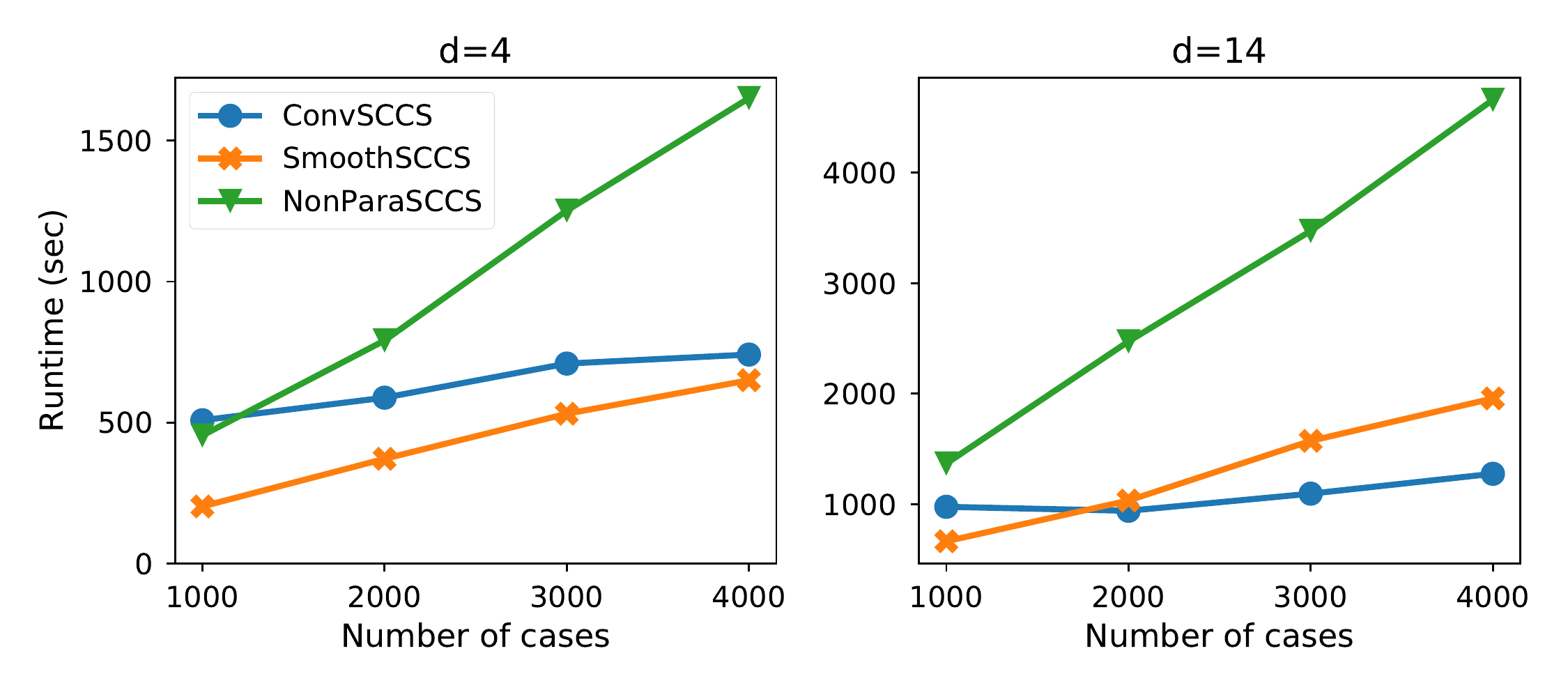}
  \caption{
  Runtimes of ConvSCCS, SmoothSCCS and NonParaSCCS described Section \ref{sec:experiments} for $1000,~2000,~3000,~4000$ cases. \emph{Left:} runtimes on $4$ features. \emph{Left:} runtimes on $14$ features.
  As SmoothSCCS and NonParaSCCS can only handle one feature at a time, we report the time required to fit them on each studied feature while ConvSCCS is fitted on all the features simultaneously.
  For each model, a fit includes cross-validation of the hyper-parameters and estimation of confidence bands. Confidence bands of ConvSCCS are estimated using parametric bootstrap, with $200$ bootstrap samples.
  }\label{runtimes}
\end{figure}

Figure~\ref{runtimes} gives the run times of all three procedures.
ConvSCCS seems to scale better than both SmoothSCCS and NonParaSCCS when fitting a large number of feature such as $d=14$ on $m > 2000$ cases.
However, in small studies, for example when $d=4$, SmoothSCCS is the fastest algorithm, while NonParaSCCS is overall slower than the two other algorithms.
According to its improved performance and scalability when studying several drugs, ConvSCCS seems to be a useful model for ADR screening on LODs.

\subsection{Application on data from the French national health insurance information system}
\label{sub:cnam-application}

The association between pioglitazone and risk of bladder cancer among diabetic patients in France was investigated by~\cite{Neumann2012} with a population-based cohort study.
The data is extracted from the French national health insurance information system (\textit{Syst\`{e}me National d'Information Inter-r\'{e}gimes de l'Assurance Maladie} ; SNIIRAM ; see~\cite{TUPPIN2010286}) linked with the French hospital discharge database (\textit{Programme de M\'{e}dicalisation des Syst\`{e}mes d'Information} ; PMSI ; see~\cite{ATIH} website).
The cohort included patients covered by the general insurance scheme aged 40 to 79 years on 2006/12/31 who filled at least one prescription for a glucose-lowering drug in 2006. The end of the observation period was set on 2009/12/31. The glucose-lowering drugs investigated were insulin, metformin, sulfonylurea, pioglitazone, rosiglitazone and other oral hypoglycemic agent. The authors defined drug exposures as time-dependent variables.
The exposure starts when the patient buys the drug two times in a 6-months window, setting the beginning of the exposure in the middle of the 6-months window. Once a patient has been exposed, he is considered as exposed until the end of his follow-up.
The cumulative dose and duration of pioglitazone were investigated as well.
Follow-up started 6 months after study entry to allow for sufficient time to observe drug exposure.
So although the depth of the data was 48 months, the cohort was followed for up to 42 months. \cite{Neumann2012} used
a multivariate Cox  model to estimate the bladder cancer hazard ratios for glucose-lowering drugs exposures, adjusted for age using groups of 5 years and gender.
A significant association between exposure to pioglitazone and an increase in the risk of bladder cancer among male patients was found.
The risk was significantly increased for long duration of exposures. As a result of this study, the use of pioglitazone was suspended in France in June 2011.

Such an approach is hypothesis driven (pioglitazone is increasing the risk of bladder cancer) and calls for ad hoc and a priori features to be built. 
The process of designing these features is time consuming, involves a work of precision done by experts of the domain and dedicated to the targeted drug.
ConvSCCS is different: it does not focus on a single drug and does not require any specific features for each drug. 
In that sense, it is much more scalable (in the number of drugs being under study).
Moreover, it allows to infer the longitudinal effects of each drug and benefits from being a SCCS model, with the advantages explained in Section~\ref{sec:model}.

In the context of a research partnership between Ecole Polytechnique and CNAMTS, we have access to the full SNIIRAM/PMSI database.
This is an SQL database containing hundreds of tables built around medical reimbursements of more than 65 million people (its size is between 150 and 200 TB). 
Our team set up a 15 nodes Spark cluster and developed en ETL (Extract Transform Load) pipeline to transform the data into a single patient-centric table that can be used to build features that feed various statistical inference algorithms. 
We used this pipeline to extract the features and reproduce the cohort described in~\cite{Neumann2012}. Demographics and drug use statistics of the resulting cohort are described in Table~\ref{cohort_stats}.

\begin{table}[!htbp]
\centering
\small
    \begin{tabular}{lr}
        \hline
        \textbf{Characteristics} & \textbf{Overall study population} \\ \hline
        N               & 1,428,637                     \\
        Men             & 771,647                       \\
        Bladder cancers & 1,804      \\
        \emph{Age (years)}     &                               \\
        40-44           & 54,989                        \\
        45-49           & 94,986                        \\
        50-54           & 160,388                       \\
        55-59           & 238,611                       \\
        60-64           & 238,394                       \\
        65-69           & 223,721                       \\
        70-74           & 232,100                       \\
        75-79           & 185,448                       \\
        \emph{Number of patients exposed to glucose-lowering drugs} & \\
         \emph{      (a patient can appear in several lines)} &      \\
				Insulin              &   343,912                \\
				Other OHA            &   434,352                \\
        Rosiglitazone        &   157,346                \\
        Metformin            & 1,043,967                \\
				Pioglitazone         &   158,619                \\
				Sulfonylurea         &   836,572                \\
				\emph{Number of patients exposed to a single glucose-lowering drug} &    \\
         \emph{      (each patient appears at most in a single line)} &      \\

				Insulin              &   102,021                \\
				Other OHA            &    34,927                \\
        Rosiglitazone        &     2,239                \\
        Metformin            &   208,331                \\
				Pioglitazone         &     4,486                \\
				Sulfonylurea         &   145,509                \\
        \hline
    \end{tabular}
    \caption{Demographics and glucose-lowering drug use of the cohort of French diabetic patients covered by the general insurance scheme (i.e., in the SNIIRAM database), aged 40-79 years and followed from 2006 to 2009.}
    \label{cohort_stats}
\end{table}


We apply ConvSCCS to the cohort of $1804$ patients with bladder cancer.
The goal of this analysis is to discover again the positive association between pioglitazone and bladder cancer in a multi-drugs setting.
We use the same definition of outcomes as in~\cite{Neumann2012}.
Since the definition of exposure considered herein might introduce a bias in the latencies estimated by our model, our definition of exposure differ slightly.
We consider that patients are exposed to a molecule as soon as they purchase a drug containing this molecule.
We use 30-days time intervals, since we don't have access to a smaller granularity.
We use calendar time and control the effect of age using $4$-years groups, and consider a risk window of 24 months.
We selected the best hyper-parameters using stratified 3-fold cross-validation, with random search.
Bootstrap confidence intervals are computed with $200$ bootstrap samples obtained with the parametric bootstrap on the unpenalized likelihood.
We refit the model in the support of the parameters obtained with the penalized procedure before using the bootstrap.

\begin{figure}[!h]
  \centering
  \includegraphics[width=\textwidth]{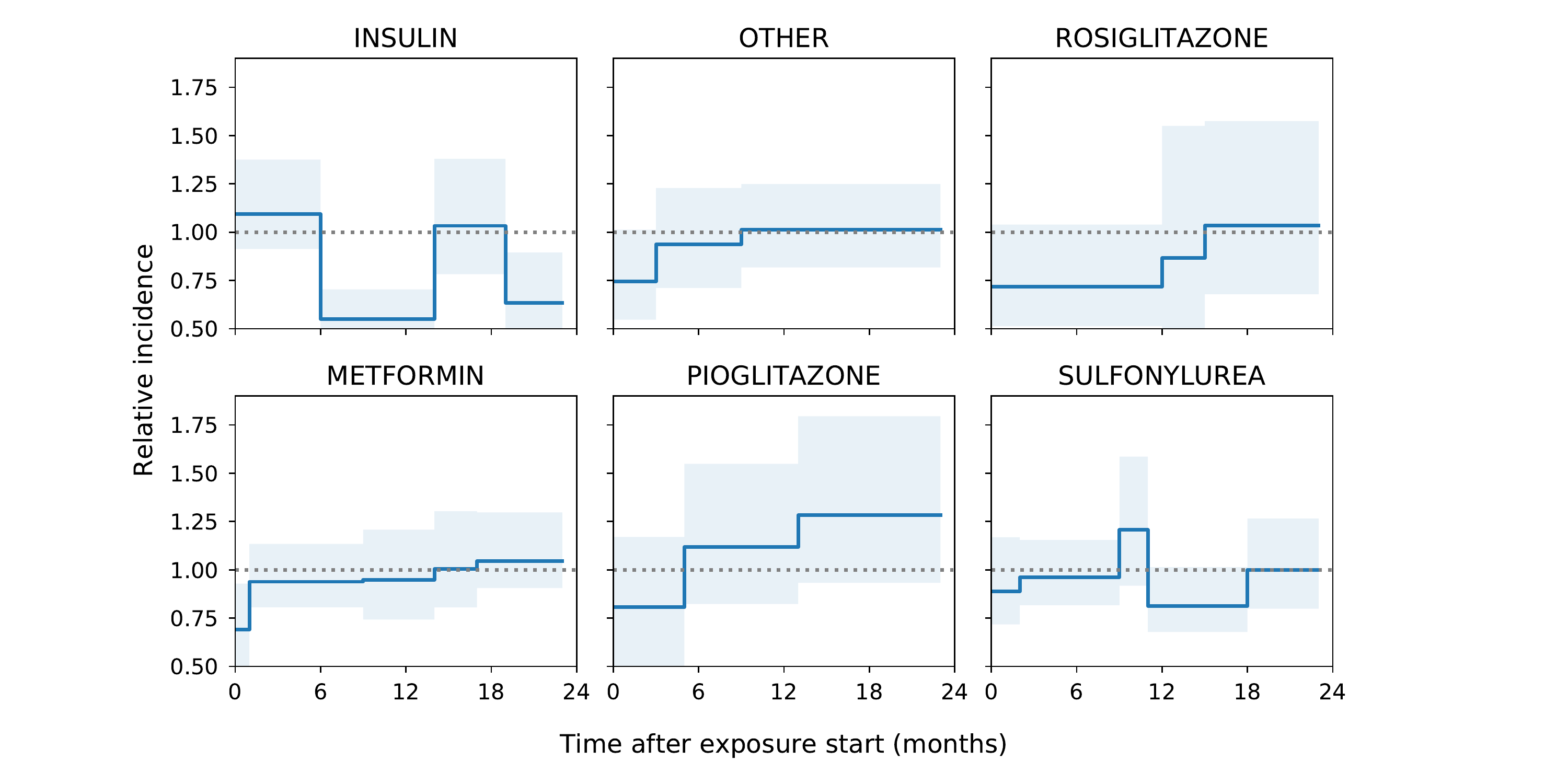}
  \caption{Estimated relative incidences of glucose lowering drugs on the risk of bladder cancer.
  Blue curves represent the estimated relative incidences $k=0, \dots, 23$ months after the beginning of exposure.
  Blue bands represent $95\%$ confidence intervals estimated by the parametric bootstrap, with $200$ bootstrap samples.}
  \label{pio_results}
\end{figure}

The estimated relative incidences and $95\%$ bootstrap confidence intervals for all investigated glucose-lowering drugs are represented in Figure~\ref{pio_results}.
Cross validation and $95\%$ bootstrap confidence intervals computation took $188$ seconds using a single thread of an Intel Xeon E5-2623 v3 3.00 Ghz CPU.
Thanks to the penalization used in ConvSCCS, the estimated relative incidences and confidence intervals are piecewise constant on large steps: this is particularly interesting, since it allows to detect only significant variations of the relative risks.
We recover a strong increase of the risk related to an exposure to pioglitazone after one year.
Indeed, Figure~\ref{pio_results} shows a positive association between pioglitazone and the risk of bladder cancer increasing over time from 5 to 24 months after exposure start, while other hypoglycemic agents do not seem to have any incidence on the risk.
Our results are therefore consistent with~\cite{Neumann2012}.
Our model estimated a hazard ratio of $0.81$ ($[0.48, 1.17]$) for pioglitazone cumulative exposure less than $5$ months and a hazard ratio of $1.12$ ($[0.82, 1.54]$) for pioglitazone cumulative exposure between $5$ and $12$ months. \cite{Neumann2012} found a hazard ratio of $1.05$ ($[0.82, 1.36]$) for pioglitazone cumulative exposure less than $12$ months. For exposure greater than $12$ months, they estimated a hazard ratio of $1.34$ ($[1.02, 1.75]$) whereas our model found $1.28$ ($[0.93, 1.79]$).

\section{Conclusion}
\label{sec:conclusion}

In this paper, we introduce ConvSCCS, a multivariate SCCS method with a flexible risk formulation and show how it can be used to perform ADR screening on LODs.
We use a discrete-time version of the SCCS model~\cite{Farrington1995}, modeling exposures-related relative incidences with low-granularity step functions.
Our model does not require a precise formulation of risk windows, as it allows the risk to vary throughout the risk window.
To avoid overfitting, we propose a penalization controlling the regularity of the step functions and providing exposure-related features selection. To our knowledge, this is the first multivariate SCCS model providing flexible formulation for the intensity and feature selection.

ConvSCCS is faster and has better performances than the state-of-the-art when studying many features, while having comparable performances when studying few features.
We applied our model on a cohort of diabetic patients studied in~\cite{Neumann2012}, and showed that we are able to recover the ADR detected by the authors using our model.
ConvSCCS can therefore be seen as a tool for exploratory analyses, providing a description of the relative risk functions for many drugs at a time.
As a result, this model can be useful to generate hypotheses, or to calibrate studies whose goal is to quantify relative risks.

\section*{Acknowledgments}

This research benefited from the CNAMTS-Polytechnique research partnership, and from the Data Science Initiative of Ecole polytechnique.

We thank Aurélie Bannay, Hélène Caillol, Jo\"el Coste, Claude Gissot, Anke Neumann, Jérémie Rudant, and Alain Weill for their insights and their help with the understanding of the SNIIRAM database.
We also thank Firas Ben Sassi, Prosper Burq, Xristos Giastidis, Sathiya Kumar, Daniel de Paula and Youcef Sebiat that are the research engineers working on this partnership.

\appendix

\section{Likelihood in SCCS models}
\label{supp:likelihood_sccs}

From~\cite{DaleyD.J.2003a}, the Poisson likelihood of a single patient $i$ can then be written as
\begin{equation*}
	L_i(n_i ; t_{i} | X_i) = e^{- \int_{a_i}^{b_i} \lambda_i(s,X_i) ds} \prod_{k=1}^{n_i}
	\lambda_i(t_{ik}, X_i),
\end{equation*}
and the total number of events $n_i = N_i([a_i,b_i])$ follows a Poisson distribution
\begin{equation*}
	\P(n_i |X_i) = \frac{(\int_{a_i}^{b_i} \lambda_i(s,X_i) ds)^{n_i}}{n_i !} e^{- \int_{a_i}^{b_i}
	\lambda_i(s,X_i) ds}.
\end{equation*}
Conditioning the likelihood by the total number of events and on the covariates histories leads to the SCCS likelihood of a patient history
\begin{align*}
L_{i}(t_{i} | n_i, X_i) &= \frac{L_{i}(n_i ; t_{i} | X_i)}{\P(n_i|X_i)} \\
&=  \frac{e^{- \int_{a_i}^{b_i} \lambda_i(s,X_i) ds} \prod_{k=1}^{n_i}
\lambda_i(t_{ik}, X_i)}{ e^{- \int_{a_i}^{b_i} \lambda_i(s,X_i) ds\frac{(\int_{a_i}^{b_i}
\lambda_i(s,X_i) ds)^{n_i}}{n_i !} }} \\
&=  n_i! \prod_{k=1}^{n_i} \frac{\lambda_i(t_{ik}, X_i)}{\int_{a_i}^{b_i} \lambda_i(s,X_i) ds},
\end{align*}
where we used the convention  $\prod_{k=1}^{0} \ldots = 1$ (i.e., the likelihood is equal to $1$ if a patient does not have any event, namely $n_i=0$).
The likelihood of $m$ patients can therefore be expressed, up to constants independent on the intensities, as
\begin{equation*}
	L \propto \prod_{i = 1}^m \prod_{k=1}^{n_i} \frac{ \lambda_i(t_{ik}, X_i)}{\int_{a_i}^{b_i}
	\lambda_i(s,X_i) ds}.
\end{equation*}

\section{Discrete time SCCS}
\label{supp:discrete_sccs}
We assume that, for $i=1, \ldots, m$, the intensity $\lambda(t, X_i(t))$ is constant over time intervals
$I_k = (t_k,t_{k+1}]$, $k=1, \dots, K$ that form a partition of the observation interval $(a, b]$.
We choose $I_k$ to be of constant length, chosen without loss of generality equal to $1$.
In practice, we use the smallest granularity allowed by data.
We therefore can assume that $(a_i, b_i]\cap I_k$ is either $\emptyset$ of $I_k$ for all $i=1, \ldots, m$, and $k=1, \dots, K$, which means that the observation period of each individual is a union of intervals $I_k$.
The discrete-time likelihood writes
\begin{align*}
L(t_i;n_i|X_i) &= \exp \Big( \sum_{k=1}^K \int_{I_k} \log(\lambda(s, X_i(s)))  \text{d} N_i(s)
- \sum_{k=1}^K \int_{I_k} \lambda(s, X_i(s))  \text{d} s \Big) \\
&=  \exp \Big( \sum_{k=1}^K \log(\lambda_{ik}) N_i(I_k) - \sum_{k=1}^K  \lambda_{ik} \Big),
\end{align*}
where $\lambda_{i, k}$ is the value of $\lambda(t, X_i(t))$ for $t \in I_k$, where
$N_i(I_k) = \int_{I_k} d N_i(t)$ and where we used $\int_{I_k} dt = 1$ and the fact that $N_i(I_k) = 0$
and $\lambda_{ik} = 0$ if $I_k \cap (a_i,b_i]=\emptyset$.
The distribution of the total number of events for patient $i$ is given by
\begin{equation*}
\P(n_i | X_i) = \frac{ \big(\int_{a_i}^{b_i} \lambda(s, X_i(s)) \text{d}s
\big)^{n_i}}{n_i !} e^{- \int_{a_i}^{b_i} \lambda(s, X(s)) \text{d}s}
=  \frac{ \big(\sum_{k=1}^K \lambda_{ik} \big) ^{n_i}}{n_i !} e^{- \sum_{k=1}^K  \lambda_{ik}},
\end{equation*}
which leads to
\begin{align*}
L(t_i | n_i, X_i) = \frac{L(n_i;t_i|X_i)}{\P(n_i | X_i)}
&= \frac{ \exp \big(\sum_{k=1}^K  \log(\lambda_{ik}) N_i(I_k) - \sum_{k=1}^K  \lambda_{ik}\big)}{
\frac{(\sum_{k=1}^K  \lambda_{ik}) ^{n_i}}{n_i !} e^{- \sum_{k=1}^K  \lambda_{ik} }} \\
&= n_i ! \prod_{k=1}^K  \Big( \frac{ \lambda_{ik}}{\sum_{k'=1}^K  \lambda_{ik'}} \Big) ^{N_i(I_k)},
\end{align*}
where we use the convention $0^0 = 1$, i.e. only the exposition period $(a_i,b_i]$ contributes to the likelihood, and since once again $N_i(I_k) = \lambda_{ik} = 0$ whenever $I_k \cap (a_i,b_i] = \emptyset$.
Then, defining $y_{ik} :=  N_i(I_k)$, the previous equation can be rewritten as
\begin{equation*}
	L(y_{i1}, \dots, y_{ik} | n_i, X_i) = n_i ! \prod_{k=1}^K \Big( \frac{ \lambda_{ik}
	}{\sum_{k'=1}^K \lambda_{ik'}} \Big) ^{y_{ik}}.
\end{equation*}
\section{Numerical implementation}
\label{supp:implementation}
We use the state-of-the-art SVRG algorithm from~\cite{Xiao2014} for the minimization of our penalized negative log-likelihoof. It is known to typically lead to faster convergence than quasi-newton algorithms, such as L-BFGS-B, see~\cite{liu1989limited}, while allowing to deal with non-smooth objectives.
Solving~\eqref{eq:penalized-log-lik} requires to compute the proximal operator (see~\cite{Bach2011} for a definition) of
$\pen(\theta)$.
This can be done very fast numerically: $\pen(\theta)$ can be separated into two separate proximal operators for total-variation and group-Lasso, see~\cite{Zhou2012}.
The proximal operator of group-Lasso is explicit and given by group soft-thresholding, see~\cite{Bach2011}, while the prox of total-variation is not, but can be computed very efficiently with the fast algorithm from~\cite{Condat2013a}.

\section{Software}
\label{supp:implementation}

Our model is implemented in the \texttt{Tick} library, see~\cite{Bacry2017}, which is a \texttt{Python} library focused
on statistical learning for time dependent systems.
It is open-source and available at \url{https://github.com/X-DataInitiative/tick}.
The implementation is done in \texttt{C++}, with a Python API, and is thoroughly documented at \url{https://x-datainitiative.github.io/tick/}.

\section{Simulations details}
\label{sub:simulation_details}

\paragraph{About the simulation of longitudinal features.}

Let us give some details on the way we simulated longitudinal features using Hawkes processes.
We sample $\mu_j$ using a uniform distribution on $[0, 5 \times 10^{-3}]$, which will produce unbalanced exposures in the simulations, and set $\alpha = 0.5$.
The diagonal entries $A_{j, j}$ are equal to $\mu_j$, and we sample $q$ non diagonal entries using a uniform distribution $[0, 5 \times 10^{-3}]$, while setting all other entries to zero. We set $d=4, ~q=8$ in the first experiment, $d=14, ~q=24$ in the second experiment.
We normalize $A$ so that its largest singular value is $0.1$, in order to ensure that the process does not generate too much events.
Simulation is achieved through the thinning technique,
see~\cite{ogata1981lewis}, and easily achieved using the \texttt{tick} library, see~\cite{Bacry2017}.
An example of simulated matrix $A$ is illustrated in Figure~\ref{hawkes_adjacency}.
Our simulation setup allows to generate realistic exposures, since it can reproduce the following phenomena that are typically observed in LODs:
\begin{itemize}
\item Depending on the drug, a patient using it has a higher probability to use it again in the future: this is quantified by the value of the diagonal entries $A_{j, j}$;
\item Some drugs are often purchased at the same time, because of the underlying medical
treatment: a patient using drug $j'$ has a higher probability to use drug $j$, which is quantified by $A_{j, j'}$;
\item Most of the patients use only a subset of all available drugs during their observation period, so several entries of $A$ are zeros.
\end{itemize}

\paragraph{About the risk profiles.}

\begin{figure}[!htb]
  \centering
  \includegraphics[width=\textwidth]{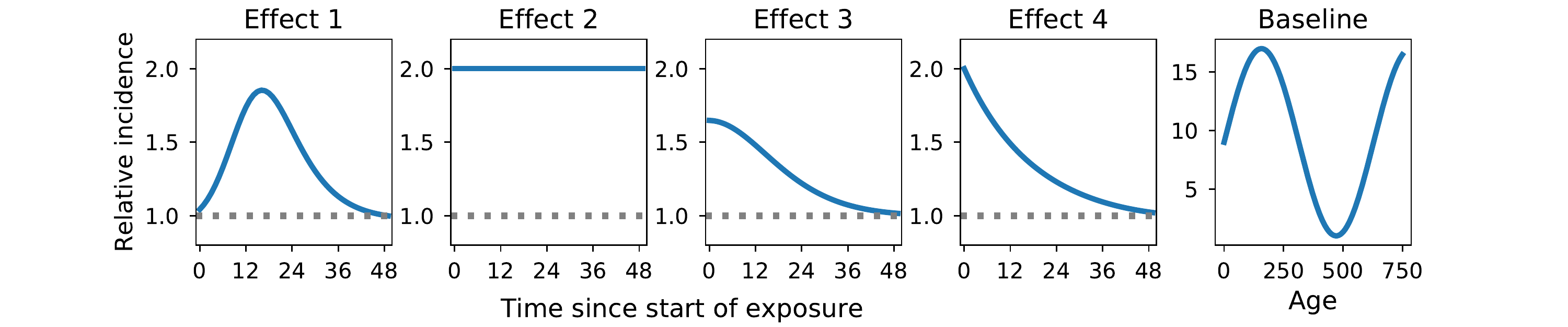}
  \caption{\emph{Left}. Set~1 of relative risk profiles.
  The effect of these relative incidences starts with the exposure, and last 50 time periods. The effect on the individual risk is multiplicative.
	\emph{Right}. Temporal baseline used in all simulations.
  }
  \label{sim_effects_1}
\end{figure}

\begin{figure}[!htb]
  \centering
  \includegraphics[width=.8\textwidth]{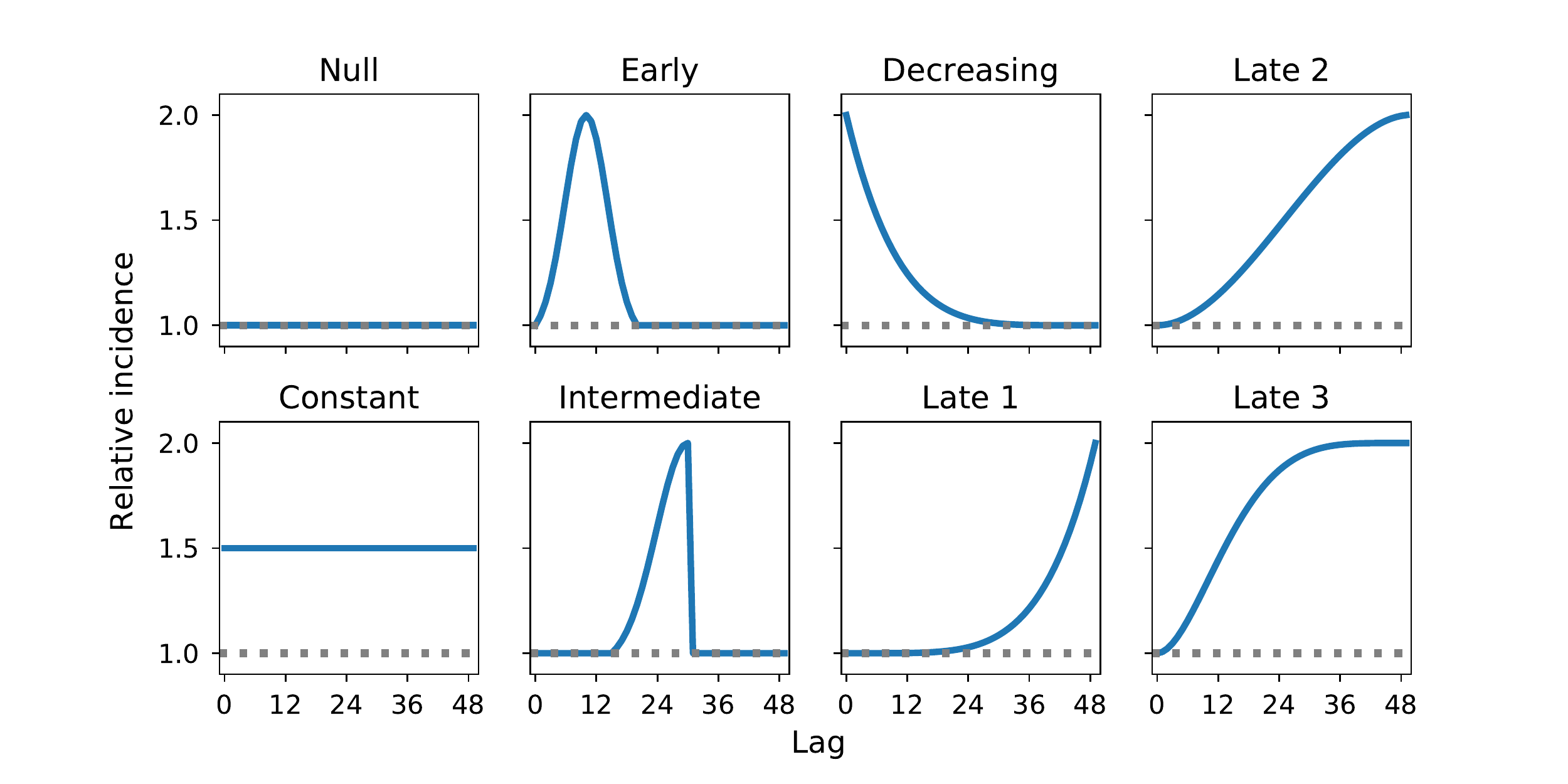}
  \caption{Set~2 of relative risk profiles. The effect of these relative incidences starts with the exposure, and last at most 50 time periods. The effect on the individual risk is multiplicative. Note that we include~$7$ features with the ``null'' risk profile in addition to one feature with each other risk profile in Set~2, to simulate datasets in which there are irrelevant features.}\label{sim_effects_2}
\end{figure}

We provide below a precise description of the two sets of risks profiles considered in our simulations.
\begin{itemize}
\item Set~1 of risk profiles corresponds to the ones used in~\cite{Ghebremichael-Weldeselassie2017}, and are represented in Figure~\ref{sim_effects_1}.
We use a lower order of magnitude than ~\cite{Ghebremichael-Weldeselassie2017}, resulting in risk profiles with maximum between $1.5$ and $2$ matching the magnitudes encountered in our application.
The first risk profile is unimodal, the second has a constant effect, two others are continuously decreasing. In this set, risk profiles length matches $p = 50$.
\item Set~2 of risk profiles represent effects described in~\cite{Aronson1222}: rapid, early, intermediate, late and delayed effect, see~Figure~\ref{sim_effects_2}, with magnitudes similar to Set~1.
It contains the four shapes from Set~1, and a null risk, a unimodal risk with a sharp drop and three shapes of continuously increasing risks. This set contains risk profiles for which the optimal risk period is smaller than $p=50$. We generate $7$ features with ``Null'' risk profile, and one feature for each other risk profile, resulting in $14$ features.
\end{itemize}
Following~\cite{Ghebremichael-Weldeselassie2017}, we use for all patients a baseline relative incidence given by $\phi(t) \propto 8\sin(.01t) + 9$ (see the right-hand side of Figure~\ref{sim_effects_1}) which can be thought as the effect of age whenever each patient has the same age.

\paragraph{About the performance measure.}

As defined in Section \ref{relative_incidence}, relative incidence of drug $j$, $k$ periods after exposure start is defined as $\hat{r}_k^j = \exp(\hat{\theta}^j_k)$, $k=0, \dots, p$ in our model.
In \cite{Ghebremichael-Weldeselassie2016,Ghebremichael-Weldeselassie2017}, the estimated relative incidence is defined as $\hat{r}_k^j = \hat{\theta}^j(k) >0$ for $k=0, \dots, p$, see Equation~\eqref{eq:gw-intensity}.
Denoting the ground truth relative incidence $r^*$, the  MAE is given by
\begin{equation*}
	MAE = \frac{1}{dK} \sum_{j=1}^d \sum_{k=1}^K | r_k^{j*} - \hat{r}_k^j |.
\end{equation*}
Since we assume that all the patients are affected by the baseline in the same way, its order of magnitude cannot be properly estimated by the models.
In order to be able to compare baseline relative risks, we constrain their integral to be equal to one as~\cite{Ghebremichael-Weldeselassie2017}.

\end{document}